\documentclass[journal]{IEEEtran}

\IEEEoverridecommandlockouts                              
\usepackage{graphics} 
\usepackage{epsfig} 
\usepackage{mathptmx} 
\usepackage{times} 
\usepackage{amsmath} 
\usepackage{amssymb}  
\usepackage{floatrow}
\usepackage{multirow}
\usepackage{algorithm}
\usepackage{algpseudocode}
\usepackage{color}

\title{\LARGE \bf
Sampling-based Learning Control for Quantum Systems with Uncertainties}
\author{Daoyi~Dong, Mohamed A. Mabrok, Ian R.~Petersen, 
Bo Qi, Chunlin~Chen, Herschel Rabitz
\thanks{This work was supported by the Australian Research Council (DP130101658 and
FL110100020), by the Natural Science Foundation of China (Grants No. 61273327, No. 61374092 and No. 61432008), the NSF (CHE-1058644) and the ARO (W911NF-13-1-0237) and ARO-MURI (W911NF-11-1-2068).}
\thanks{D. Dong, M. A. Mabrok and I. R. Petersen are with the School of Engineering and Information Technology, University of New South Wales, Canberra, ACT 2600, Australia (email:
daoyidong@gmail.com; M.Mabrok@adfa.edu.au; i.r.petersen@gmail.com).}
\thanks{B. Qi is with the Key Laboratory of Systems and Control,
Academy of Mathematics and Systems Science, Chinese Academy
of Sciences, Beijing 100190, China (email: qibo@amss.ac.cn).}
\thanks{C. Chen is with the Department of Control and System Engineering, School of Management and Engineering, Nanjing
University, Nanjing 210093, China (Email: clchen@nju.edu.cn).}
\thanks{H. Rabitz is with the Department of
Chemistry, Princeton University, Princeton, New Jersey 08544, USA
(email: hrabitz@princeton.edu).}}

\begin{document}

\maketitle
\begin{abstract}
Robust control design for quantum systems has been recognized as a key task in the development of practical quantum technology. In this
paper, we present a systematic numerical methodology of sampling-based
learning control (SLC) for control design of quantum systems with uncertainties.
The SLC method includes two steps of ``training" and ``testing". In the training step, an augmented system is
constructed using artificial samples generated by sampling uncertainty parameters according to a given distribution. A gradient flow based
learning algorithm is developed to find the control
for the augmented system. In the process of testing, a number of additional samples are tested to
evaluate the control performance where these samples are obtained through sampling the uncertainty parameters according to a possible distribution. The SLC method is applied to three significant examples of quantum robust control including state preparation in a three-level quantum system, robust entanglement generation in a two-qubit superconducting circuit and quantum entanglement control in a two-atom system interacting with a quantized field in a cavity. Numerical results
demonstrate the effectiveness of the SLC
approach even when uncertainties are quite large, and show its potential for robust control design of
quantum systems.
\end{abstract}


\section{Introduction}\label{Sec1}
The control and manipulation of quantum phenomena lie at the heart of developing practical quantum technologies, and the exploration of quantum
control theory and methods is drawing wide interests from scientists and engineers
\cite{Dong and Petersen 2010IET}-\cite{Brif et al 2010}. In the development of practical quantum technologies, robustness has been recognized as a key performance measure
since the existence of uncertainties and noises is unavoidable in the modeling and control process for real quantum
systems \cite{Pravia et al 2003}-\cite{James 2004}. For example, the chemical shift may not be exactly known in the model of a spin system in nuclear magnetic resonance (NMR) \cite{Khaneja 2005}, \cite{Li and Khaneja 2009}. In a superconducting quantum circuit, there exist possible fluctuations in the coupling energy of a Josephson junction \cite{Makhlin et al 2001}, \cite{Montangero et al 2007}. In the dipole approximation for a molecular system interacting with laser fields, imprecision in the model parameters is unavoidable \cite{Zhang and Rabitz 1994}. It is also common that there exist errors in control pulses or fields applied to quantum systems. Hence, it is important to develop systematic robust design methods for the analysis and synthesis of quantum systems with uncertainties. Several methods have been
proposed for the robust control of quantum systems \cite{James et al 2007}-\cite{Yamamoto and Bouten 2009}. For example,
James \emph{et al}. \cite{James et al 2007} have formulated and
solved an $H^{\infty}$ controller synthesis problem for a class of
quantum linear stochastic systems. Adiabatic techniques (e.g., STIRAP - stimulated Raman adiabatic passage) \cite{Vasilev PRA et al 2009}-\cite{Boscain et al 2012} have been widely applied to robust control problems of quantum systems when the adiabatic limit can be satisfied. Optimized composite pulses have been applied in NMR to improve robustness performance \cite{Khaneja 2005}, \cite{Owrutsky and Khaneja 2012}. A noise filtering approach has been presented to enhance robustness in quantum control \cite{Biercuk et al NaturePhy 2014}. A sequential convex programming method has been proposed for designing robust quantum gates \cite{Kosut et al 2013}. A sliding mode control
approach has been presented to deal with Hamiltonian uncertainties in two-level quantum systems \cite{Dong and Petersen 2009NJP}, \cite{Dong and Petersen 2011Automatica}.

In classical (non-quantum) control systems, feedback control is the dominant method for robust control design. Feedback control (including measurement-based feedback control and coherent feedback control) has been applied to some quantum systems for improved control performance \cite{Wiseman and Milburn 2009}, \cite{Sayrin et al 2011}-\cite{Mirrahimi and van Handel 2007}. However, open-loop control is more practical than feedback control for most quantum systems with current technology considering the small time scales and measurement backaction in the quantum domain \cite{Kosut et al 2013}. Several open-loop control strategies have been presented to design robust control laws for specific quantum systems. For example, dynamical decoupling has been developed for control design of quantum systems with uncertainties \cite{Viola et al 1999}-\cite{Biercuk et al 2009}. Existing results showed that control fields designed by learning have the property of robustness \cite{Kosut et al 2013}, \cite{Rabitz et al 2000}, \cite{Chakrabarti and Rabitz 2007}. Recently, Zhang \emph{et al.} \cite{Zhang et al 2012} employed a gradient-based algorithm to design robust control pulses for electron shuttling. They considered that parameter uncertainties exist in the energy difference when an electron is transported along a chain of donors. An effective optimization algorithm has been developed to find robust control pulses by discretizing the uncertainty range and deriving the gradient of an aggregate fidelity with respect to sinusoidal control fields \cite{Zhang et al 2012}.

In this paper, we present a systematic numerical methodology of sampling-based learning control (SLC) for robust
design of quantum systems with uncertainties. Sampling-based learning control was first presented for control design of inhomogeneous quantum ensembles where the SLC method includes two steps of ``training" and ``testing" \cite{Chen et al 2013arXiv}. A generalized system is constructed from some samples with different values of the inhomogeneous parameters in the training step and a control field is learned through a gradient flow based optimization algorithm. The control is evaluated using additional samples for some possible values of the inhomogeneous parameters. The results showed that the SLC approach can find an effective control law to drive the members in an inhomogeneous ensemble to a given target state with high fidelity.
In this paper, we contribute a systematic SLC method with specific learning algorithms for the robust control of quantum systems with uncertainties. In particular, we generate artificial samples by sampling the uncertainty parameters in the system model and construct an
augmented system using these samples in the training step \cite{Dong et al 2013CDC}. Then a gradient
flow based learning and optimization algorithm is developed to learn a control law
with desired performance for the augmented system. In the
process of testing, we test a number of
samples of the uncertainties to evaluate the control performance. The SLC method is applied to three significant examples of quantum robust control. The first example is a three-level quantum system that is found widely in natural atoms and artificial atoms \cite{You and Nori 2011}. A problem of state preparation is investigated when uncertainties exist in the three-level system. In the second example, we consider superconducting quantum circuits which have been recognized as promising candidates for quantum information processing due to their advantages of scalability and design flexibility (see, e.g., \cite{Clarke and Wilhelm 2008}, \cite{You and Nori 2005}, \cite{Xiang et al 2013}, \cite{Dong et al 2015SRep}). In particular, we employ the SLC method to learn a robust control law that can be applied to a two-qubit superconducting circuit for generating quantum entanglement. The third example investigates the application of the SLC approach to the robust control of quantum entanglement in a two-atom system interacting with a quantized field in a cavity \cite{Altomare2010a}, \cite{Mabrok et al 2014CDC}. Numerical
results show that the SLC method is effective for
robust control design of these classes of quantum systems with uncertainties.

This paper is organized as follows. Section \ref{Sec2} formulates
the quantum control problem. Section \ref{Sec3}  presents the
sampling-based learning control approach and introduces  a gradient
flow based learning algorithm. Numerical results on
control design in three-level quantum systems are
presented in Section \ref{Sec4}. The SLC method is applied to robust entanglement generation in a two-qubit superconducting circuit in Section \ref{Sec5}. In Section \ref{Sec6}, the SLC approach is used to learn a robust control law for a two-atom system interacting with a quantized field in a cavity. Concluding remarks are
presented in Section \ref{Sec7}.

\section{Problem formulation of quantum robust control}\label{Sec2}
We focus on a finite-dimensional quantum system that can be approximated as a closed system and whose state is described using a complex vector $|\psi\rangle$ or a density operator $\rho=|\psi\rangle \langle \psi|$ in an underlying Hilbert space. The
evolution of its state $|\psi(t)\rangle$ can be described by the
following Schr\"{o}dinger equation:
\begin{equation} \label{systemmodel}
\left\{ \begin{array}{l}
  \frac{d}{dt}|{\psi}(t)\rangle=-\frac{i}{\hbar}H(t)|\psi(t)\rangle \\
 \ |\psi(0)\rangle=|\psi_{0}\rangle.\\
\end{array}
\right.
\end{equation}
The dynamics of the system are governed by a
time-dependent Hamiltonian of the form \cite{D'Alessandro 2007}
\begin{equation}\label{Hamiltonian}
H(t)=H_{0}+H_{c}(t)=H_{0}+\sum_{m=1}^{M}u_{m}(t)H_{m},
\end{equation}
where $H_{0}$ is the free Hamiltonian of the system,
$H_{c}(t)=\sum_{m=1}^{M}u_{m}(t)H_{m}$ is the time-dependent control
Hamiltonian that represents the interaction of the system with the
external fields $u_{m}(t)$ (scalar functions), and the $H_{m}$ are Hermitian operators through which external
controls couple to the system.

The solution of (\ref{systemmodel}) is given
by $\displaystyle |\psi(t)\rangle=U(t)|\psi_{0}\rangle$, where the
propagator $U(t)$ satisfies
\begin{equation}
\left\{ \begin{array}{c}
  \frac{d}{dt}U(t)=-\frac{i}{\hbar}H(t)U(t),\\
  \ U(0)=I.\\
\end{array}
\right.
\end{equation}

For an ideal model, there exist no uncertainties in (\ref{Hamiltonian}). However, for a practical quantum system, the existence of uncertainties is unavoidable due to external disturbances, imprecise models and errors in control fields. In this paper, we suppose that the system Hamiltonian has the following form
\begin{equation}\label{uncerHamiltonian}
H_{\Theta}(t)=f_{0}(\theta_{0})H_{0}+\sum_{m=1}^{M}f_{m}(\theta_{m})u_{m}(t)H_{m}.
\end{equation}
We have denoted $\Theta=(\theta_{0},\theta_{1}, \dots, \theta_{M})$ and the functions $f_{j}(\theta_{j})$ ($j=0,1,\dots, M$) characterize possible uncertainties. For example, $f_{0}(\theta_{0})$ corresponds to uncertainties in the free Hamiltonian (e.g., due to chemical shift in NMR). $f_{m}(\theta_{m})$ can characterize possible multiplicative noises in the control fields or imprecise parameters in the dipole approximation. When the $f_{m}(\theta_{m})$ are allowed to be time-dependent, the corresponding uncertainties may originate from time-varying errors in the control fields. For example, the time-dependent non-Markovian noise in the control field considered in \cite{Biercuk et al PRA 2014} can be described using the model. It is also straightforward to include additive noises in control fields by slightly modifying (\ref{uncerHamiltonian}). We assume that $f_{j}(\theta_{j})$ are continuous functions of $\theta_{j}$  and the parameters
$\theta_{j}$ could be time-dependent and $\theta_{j}\in [1-E_{j}, 1+E_{j}]$. For simplicity, we assume the uncertainty bounds $E_{0}=\dots=E_{j}=\dots=E_{M}=E$ are all equal in this paper. We assume that the nominal values of $\theta_{j}$ are 1 and the fluctuations of the uncertainty parameters $\theta_{j}$ are $2E$ (where
$E \in [0,1]$). The objective is to design the controls
$\{u_{m}(t), m=1,2,\ldots , M\}$ to steer the
quantum system with uncertainties from an initial state $|\psi_{0}\rangle$ to a target
state $|\psi_{\text{target}}\rangle$ with high fidelity. The fidelity between two quantum states $|\psi_{1}\rangle$ and $|\psi_{2}\rangle$ is defined as \cite{Nielsen and Chuang 2000}, \cite{Chen et al 2012TNNLS}:
\begin{equation}
F(|\psi_{1}\rangle, |\psi_{2}\rangle)=|\langle\psi_{1}|\psi_{2}\rangle|.
\end{equation}
The control performance is described by a \emph{performance
function} $J(u)$ for each control strategy $u=\{u_{m}(t),
m=1,2,\ldots , M\}$. The control problem can then be formulated as
a maximization problem as follows:
\begin{equation}\label{ensemble control}
\begin{split}
\displaystyle \max_u \ \ & J(u):=\max_u\mathbb{E}\vert
\langle\psi(T)|\psi_{\text{target}}\rangle\vert^{2}\\
\text{s.t.} \ \ & \frac{d}{dt}|\psi(t)\rangle=-\frac{i}{\hbar}H_{\Theta}(t)|\psi(t)\rangle, \ |\psi(0)\rangle=|\psi_{0}\rangle \\
& H_{\Theta}(t)=f_{0}(\theta_{0})H_{0}+\sum_{m=1}^{M}f_{m}(\theta_{m})u_{m}(t)H_{m},\\
& \textrm{ with } \theta_{j} \in [1-E,1+E],~ t \in [0, T].
\end{split}
\end{equation}
Note that $J(u)$ depends implicitly on the control $u$ through the
Schr\"odinger equation and $\mathbb{E}(\cdot)$ denotes the
expectation with respect to the uncertainty parameters $\Theta=(\theta_{0},\theta_{1}, \dots, \theta_{M})$.

\section{Sampling-based learning control method}\label{Sec3}
Gradient-based methods \cite{Brif et al 2010}, \cite{Long and Rabitz
2011}, \cite{Roslund and Rabitz 2009}
have been successfully applied to search for optimal solutions to a
variety of quantum control problems, including theoretical and
laboratory applications. In this paper, a gradient-based learning
method is employed to optimize the control fields for quantum systems with uncertainties. However, it is impossible to directly calculate
the derivative of $J(u)$ since there exist uncertainties and some parameters in the model are unknown. Hence,
we present a systematic numerical methodology of sampling-based learning control which includes two steps of
``training" and ``testing". In the training step, some artificial samples are generated through sampling the uncertainty parameters to construct an augmented system and a learning algorithm is developed to find an optimal control strategy for the augmented system.
Then the designed control law is applied to additional samples to test and
evaluate the control performance in the testing step.

\subsection{Sampling-based learning control}
In the training step, we first generate $N$ samples through
sampling the uncertainty parameters according to a given probability distribution
(e.g., the uniform distribution). We could choose any of the
combination $(\theta_{0n_{0}}, \theta_{1n_{1}}, \dots,
\theta_{Mn_{M}})$ of the sampled parameters $\{ \theta_{j},
j=0,1,\ldots,M\}$, where $\theta_{jn_{j}}$ is a possible value of
the uncertainty parameter $\theta_{j}$, $n_{j}=1, \dots, N_{j}$
and $N_{j}$ is the number of samples of the parameter $\theta_{j}$
($j=0, 1,\ldots,M$). The total number of potential samples is
$N=\prod_{j=0}^{M}N_{j}$.  We denote these different sample
systems as $\{n\}\ (n=1,2,\ldots,N)$ and $\Theta_{n}\in
\{(\theta_{0n_{0}}, \theta_{1n_{1}}, \dots,
\theta_{Mn_{M}})|n_{j}=1, \dots, N_{j}\}$. With these samples, we
can construct an augmented system as follows
\begin{equation}\label{augmented-system}
\frac{d}{dt}\left(%
\begin{array}{c}
  |{\psi}_{1}(t)\rangle \\
  |{\psi}_{2}(t)\rangle \\
  \vdots \\
  |{\psi}_{N}(t)\rangle \\
\end{array}%
\right)
=-\frac{i}{\hbar}\left(%
\begin{array}{c}
  H_{\Theta_1}(t)|\psi_{1}(t)\rangle \\
  H_{\Theta_2}(t)|\psi_{2}(t)\rangle \\
  \vdots \\
  H_{\Theta_N}(t)|\psi_{N}(t)\rangle \\
\end{array}%
\right),
\end{equation}
where
$H_{\Theta_n}(t)=f_{0}(\theta_{0n_{0}})H_{0}+\sum_{m}f_{m}(\theta_{mn_{m}})u_{m}(t)H_{m}$
with $n=1,2,\dots,N$. The performance function for the augmented
system is defined by
\begin{equation}\label{eq:cost}
J_N(u):=\frac{1}{N}\sum_{n=1}^N J_{\Theta_n}(u)=\frac{1}{N}\sum_{n=1}^{N}\vert \langle\psi_{n}(T)|\psi_{\text{target}}\rangle\vert^{2}.
\end{equation} The task of the training step is to find a control
strategy $u^*$
that maximizes the performance function defined in Eq.
\eqref{eq:cost}.
We will develop a gradient flow based learning algorithm to
solve this optimization problem in an iterative way \cite{Chen et al 2013arXiv}, \cite{Dong et al 2013CDC}.
Assume that the performance function is $J_N(u^{0})$ with an initial
control strategy $u^{0}=\{u^{0}_{m}(t)\}$. We can apply the
gradient flow method to obtain an (approximate) optimal control strategy
$u^{*}=\{u^{*}_{m}(t)\}$. It is clear that we may take $u^{*}$ as an approximate optimal control solution when $J_N(u)\rightarrow 1$. For the nominal system (without uncertainties), the quantum control landscape theory \cite{Chakrabarti and Rabitz 2007} has shown that there are no local maxima in the optimization problem for closed quantum systems when they are controllable and the critical points of their control landscapes are regular (for details, see, e.g., \cite{Brif et al 2010}, \cite{Chakrabarti and Rabitz 2007}). For the augmented system, it may be also possible to use a similar method to the quantum landscape theory to prove the optimal characteristics. Numerical results in this paper show that a gradient method can be used to achieve excellent performance in finding an approximate optimal solution. The detailed gradient flow algorithm
will be presented in Subsection \ref{sec2.3}.

As for the issue of choosing $N$ samples,
we generally choose them according to possible distributions of the
uncertainty parameters
$\theta_{j} \in [1-E, 1+E]$. The basic
motivation of the proposed sampling-based approach is to design the
control law using some artificial samples instead of unknown uncertainties. Therefore, it is
necessary to choose the set of samples that are representatives of
these uncertainty parameters.

For example, we consider the case with two uncertainty parameters $\theta_{0}$ and $\theta_{1}$. If the distributions of both $\theta_{0}$ and $\theta_{1}$
are uniform, we may choose equally spaced samples for
$\theta_{0}$ and $\theta_{1}$. For example, the intervals $[1-E,
1+E]$ for $\theta_{0}$ and $[1-E, 1+E]$ for $\theta_{1}$ are divided into
$N_{0}+1$ and $N_{1}+1$ subintervals, respectively,
where $N_{0}$ and $N_{1}$ are usually positive odd
numbers. Then the number of samples is $N=N_{0}N_{1}$,
and $\Theta_{n}=(\theta_{0n_{0}}, \theta_{1n_{1}})$ is chosen from the set of sample points
\begin{eqnarray}\label{discrete}
\Theta_{n} \in
\{(\theta_{0n_{0}}, \theta_{1n_{1}}): & \theta_{0n_{0}}=1-E+\frac{(2n_{0}-1)E}{N_{0}},\nonumber \ \ \ \ \ \ \ \ \ \ \ \ \ \\
\ & \theta_{1n_{1}}=1-E+\frac{(2n_{1}-1)E}{N_{1}}),\nonumber \ \ \ \ \ \ \ \ \ \ \ \ \\
\ & n_{0}\in\{1,\ldots, N_{0}\},\ n_{1}\in\{1,\ldots, N_{1}\}\}.
\end{eqnarray}

In practical applications, the numbers of $N_{0}$ and
$N_{1}$ can be chosen by experience or through
numerical computation. As long as the augmented system can model
the quantum system with uncertainties and is effective to find the optimal control
strategy, we prefer to choose small numbers for $N_{0}$ and
$N_{1}$ to speed up the training process and simplify the
augmented system. Numerical results show that five or seven samples for each uncertainty parameter are enough to achieve excellent performance.

In the testing step, we apply the optimized
control $u^{*}$ obtained in the training step to a large number of additional
samples obtained through randomly sampling the uncertainty parameters. The control performance
is evaluated for each sample in terms of
the fidelity $F(|\psi(T)\rangle,|\psi_{\text{target}}\rangle)$ between the final state
$|\psi(T)\rangle$ and the target state
$|\psi_{\text{target}}\rangle$.
If the fidelity for all the
tested samples is satisfactory, we accept the designed control
law and end the control design process. Otherwise, we go
back to the training step to find another optimized control
strategy by changing the settings (e.g., restarting the training step with a new initial
control strategy or a new set of samples).

\subsection{Gradient flow based learning algorithm}
\label{sec2.3}
To find an optimal control strategy $u^{*}=\{u^{*}_{m}(t), (t \in
[0,T]), m=1,2,\ldots, M\}$ for the augmented system
(\ref{augmented-system}), a good choice is to follow the
direction of the gradient of $J_N(u)$ as an ascent direction so as to speed up the learning process. For
ease of notation, we present the method for the case $M=1$. We introduce a
time-like variable $s$ to characterize different control
strategies $u^{(s)}(t)$. Then the gradient flow in the control space
can be defined as
\begin{equation}\label{gradientflowequation}
\frac{du^{(s)}}{ds} =\nabla J_N(u^{(s)}),
\end{equation}
where $\nabla J_N(u)$ denotes the gradient of $J_N$ with respect
to the control $u$. If $u^{(s)}$ is the
solution of \eqref{gradientflowequation} starting from an
arbitrary initial condition $u^{(0)}$, then the value of $J_N$ is
increasing along $u^{(s)}$, i.e., $\frac{d}{ds}J_N(u^{(s)})\geq
0$. In other words, starting from an initial guess $u^{0}$, we
solve the following initial value problem
\begin{equation}\label{gradientflowequation2}
\left\{%
\begin{split}
  & \frac{du^{(s)}}{ds} = \nabla J_N(u^{(s)}) \\
  & u^{(0)}=u^{0} \\
\end{split}%
\right.
\end{equation}
in order to find a control strategy which maximizes $J_N$. This
initial value problem can be solved numerically by using a forward
Euler method over
the $s$-domain, i.e.,
\begin{equation}\label{iteration1}
u(s+\triangle s, t)=u(s,t)+\triangle s\nabla J_N(u^{(s)}).
\end{equation}

For practical applications, we present an iterative
approximation version of the above algorithm to find the optimal controls $u^*(t)$ in an
iterative learning way, where we use $k$ as an index of iterations
instead of the variable $s$ and denote the control at iteration
step $k$ as $u^{k}(t)$.
Equation \eqref{iteration1} can be rewritten as
\begin{equation}\label{iteration2}
u^{k+1}(t)=u^{k}(t)+ \eta_{k}\nabla J_N(u^{k}),
\end{equation}
where $\eta_{k}$ is the updating step (learning rate) for the $k$th iteration. Using \eqref{eq:cost}, we also have
\begin{equation}
\nabla J_N(u)=\frac{1}{N}\sum_{n=1}^{N}\nabla J_{\Theta_n}(u).
\end{equation}
Recall that $J_{\Theta}(u)=\vert \langle\psi_{\Theta}(T)\vert\psi_{\textrm{target}}\rangle\vert^2$ and $\vert\psi_{\Theta}(\cdot)\rangle$ satisfies
\begin{equation}\label{app-eq:sch}
\frac{d}{dt}\vert\psi_{\Theta}\rangle=-\frac{i}{\hbar}H_{\Theta}(t)\vert\psi_{\Theta}\rangle,\quad \vert\psi_{\Theta}(0)\rangle=\vert\psi_{0}\rangle.
\end{equation}
We now derive an expression for the gradient of $J_{\Theta}(u)$ with respect to the control $u$ by using a first order perturbation. Let $\delta\psi(t)$ be the modification of $\vert \psi(t)\rangle$ induced by a perturbation of the control from $u(t)$ to $u(t)+\delta u(t)$. By keeping only the first order terms, we obtain the equation satisfied by $\delta\psi$:
\begin{eqnarray*}
\begin{split}
\frac{d}{dt}\delta\psi= & -\frac{i}{\hbar}\left(f_{0}(\theta_{0})H_0+u(t)f_{1}(\theta_{1})H_1\right)\delta\psi \\
& -\frac{i}{\hbar}\delta u(t)f_{1}(\theta_{1})H_1\vert\psi_{\Theta}(t)\rangle, \\
 \delta\psi(0)= & 0.
\end{split}
\end{eqnarray*}

Let $U_{\Theta}(t)$ be the propagator corresponding to \eqref{app-eq:sch}. Then, $U_{\Theta}(t)$ satisfies
$$\frac{d}{dt}U_{\Theta}(t)=-\frac{i}{\hbar}H_{\Theta}(t)U_{\Theta}(t),\quad U(0)=I.$$
Therefore,
\begin{eqnarray}
\delta\psi(T)=-\frac{i}{\hbar}U_{\Theta}(T)\int_0^T\delta u(t)U_{\Theta}^\dagger(t)f_{1}(\theta_{1})H_1\vert\psi_{\Theta}(t)\rangle dt\nonumber \ \ \ \ \ \\
\ =-\frac{i}{\hbar}U_{\Theta}(T)\int_0^TU_{\Theta}^\dagger(t)f_{1}(\theta_{1})H_1U_{\Theta}(t) \delta u(t)dt~ \vert\psi_0\rangle.\label{app-eq:deltapsi}
\end{eqnarray}
Using \eqref{app-eq:deltapsi}, we compute $J_{\Theta}(u+\delta u)$ as follows
\begin{eqnarray}
& J_{\Theta}(u+\delta u)-J_{\Theta}(u) \nonumber \ \ \ \ \ \ \ \ \ \ \ \ \ \ \ \ \ \ \ \ \ \ \ \ \ \ \ \ \ \ \ \ \ \ \ \ \ \ \\
 \approx&
2\Re\left(\langle\psi_{\Theta}(T)\vert\psi_{\textrm{target}}\rangle\langle\psi_{\textrm{target}}\vert\delta\psi(T)\right)\nonumber \ \ \ \ \  \ \ \  \ \ \  \ \ \ \ \ \ \ \ \\
=&2\Re\left(-i\langle\psi_{\Theta}(T)\vert\psi_{\textrm{target}}\rangle\langle\psi_{\textrm{target}}\vert \int_0^TV(t) \delta u(t)dt~ \vert\psi_0\rangle\right)\nonumber\\
=&\int_0^T2\Im\left(\langle\psi_{\Theta}(T)\vert\psi_{\textrm{target}}\rangle\langle\psi_{\textrm{target}}\vert V(t)\vert\psi_0\rangle\right)\delta u(t)dt,  \ \ \ \label{app-eq:dJ}
\end{eqnarray}
where $\Re(\cdot)$ and $\Im(\cdot)$ denote, respectively, the real and imaginary parts of a complex number, and $V(t)=U_{\Theta}(T)U_{\Theta}^\dagger(t)f_{1}(\theta_{1})H_1U_{\Theta}(t)$.

Recall also that the definition of the gradient implies that
\begin{eqnarray}
J_{\Theta}(u+\delta u)-J_{\Theta}(u)=&\langle \nabla J_{\Theta}(u),\delta u\rangle_{L^2([0,T])}+o(\Vert\delta u\Vert)\nonumber\\
=&\int_0^T \nabla J_{\Theta}(u)\delta u(t)dt+o(\Vert\delta u\Vert).\label{app-eq:dJ2}
\end{eqnarray}Therefore, by identifying \eqref{app-eq:dJ} with \eqref{app-eq:dJ2}, we obtain
\begin{equation}\label{app-eq:gradJ}
 \nabla J_{\Theta}(u)=2\Im\left(\langle\psi_{\Theta}(T)\vert\psi_{\textrm{target}}\rangle\langle\psi_{\textrm{target}}\vert V(t)\vert\psi_0\rangle\right).
\end{equation}
The gradient flow method can be generalized to the case with $M>1$
as shown in \emph{Algorithm 1}. For a
termination criterion of the iterative learning process, we use the following: if the
change of the performance function for $100$ consecutive iterations is less than a given
small threshold $\epsilon>0$, i.e.,
$|J(u^{k+100})-J(u^{k})|<\epsilon$, we end the
learning process. In this paper we choose $\epsilon=10^{-4}$ for all
numerical experiments.

\begin{algorithm}

\caption{Gradient flow based iterative learning}
\label{ModifiedGradientFlow}

\begin{algorithmic}[1]

\State Set the index of iterations $k=0$

\State Choose a set of arbitrary controls $u^{k=0}=\{u_{m}^{0}(t),\
m=1,2,\ldots,M\}, t \in [0,T]$

\Repeat {\ (for each iterative process)}

\Repeat {\ (for each training sample
$n=1,2,\ldots,N$)}

\State Compute the propagator $U_{\Theta_n}^{k}(t)$ with the control
strategy $u^{k}(t)$


\Until {\ $n=N$}

\Repeat {\ (for each control $u_{m}\ (m=1,2,\ldots,M)$ of the control
vector $u$)}

\State
$\delta_m^{k}(t)=\frac{2}{N}\sum^N_{n=1}\Im\left(\langle\psi_{n}(T)\vert\rho_{\textrm{target}}
V^k_{\Theta_n}(t)\vert\psi_0\rangle\right)$ where
$\rho_{\textrm{target}}=\vert\psi_{\textrm{target}}\rangle\langle\psi_{\textrm{target}}\vert$,
$V^k_{\Theta_n}(t)=U^k_{\Theta_n}(T)(U_{\Theta_n}^k(t))^{\dagger}f_{m}(\theta_{mn_{m}})H_mU^k_{\Theta_n}(t)$
and $n_{m}\in \{1, 2, \dots, N_{m}\}$

\State $u_{m}^{k+1}(t)=u_{m}^{k}(t)+\eta_{k} \delta_{m}^{k}(t)$

\Until {\ $m=M$}

\State $k=k+1$

\Until {\ the learning process ends}

\State The optimal control strategy
$u^{*}=\{u_{m}^*\}=\{u_{m}^{k}\}, \ m=1,2,\ldots,M$

\end{algorithmic}
\end{algorithm}

In practical applications, it is usually difficult to find the
numerical solution to a time varying continuous control strategy
$u(t)$ using \emph{Algorithm 1}. In simulation, we usually divide
the time interval $[0,T]$ equally into a number of smaller time
intervals $\triangle t$ and assume that the controls are constant
within each $\triangle t$. Instead of $t \in [0,T]$, the time
index will be $t_{w}=wT/W$, where $W=T/\triangle t$ and
$w=0,1,\ldots,W$.

In the following three sections, we apply the SLC method to three examples. The first example is the state preparation in a general three-level quantum system where the main focus is on demonstrating the SLC method. We assume that there are no constraints on the control fields and the uncertainty parameters have uniform distributions or time-varying distributions. In the second example, we consider entanglement generation in superconducting quantum circuits. We use some practical models and relevant parameters in the literature. Bounded control fields and truncated Gaussian distributions for uncertainty parameters are assumed. The third example considers the application of the SLC method to a two-atom system interacting with a quantized field.

\section{State preparation in three-level quantum systems}\label{Sec4}
In this section, we demonstrate the application of the proposed
SLC method to robust state preparation in a $V$-type three-level
quantum system with uncertainties. $V$-type three-level systems are a typical class of quantum systems in atomic physics. Some natural and artificial atoms can be described by a $V$-type three-level model \cite{You and Nori 2011}. State preparation is an essential task in quantum information processing \cite{Nielsen and Chuang 2000}. For example, specific quantum states are required to be prepared for initialization in quantum computation and transfer in quantum communication. It is important to achieve robust preparation of these specific states for practical applications of quantum technology. For simplicity, we assume no constraints on the external controls and use atomic units (i.e., setting $\hbar=1$) in this section. The aim is to show how to apply the proposed SLC method for robust control design of quantum systems with uncertainties.

\subsection{State preparation in a $V$-type quantum system}
We consider a $V$-type quantum system and
demonstrate the SLC design process. Assume that the initial state is
$|\psi(t)\rangle=c_{1}(t)|1\rangle+c_{2}(t)|2\rangle+c_{3}(t)|3\rangle$.
Let $C(t)=(c_{1}(t),c_{2}(t),c_{3}(t))$ where the $c_i(t)$'s are complex numbers. We have
\begin{equation}
i\dot{C}(t)=(f_{0}(\theta_{0})H_{0}+\sum_{m=1}^{M}f_{m}(\theta_{m})u_{m}(t)H_{m})C(t).
\end{equation}
We take $H_{0}=\textrm{diag}(1.5, 1, 0)$ and choose $H_{1}$, $H_{2}$,
$H_{3}$ and $H_{4}$ as follows \cite{Hou et al 2012}:
\begin{equation}\label{h0}
H_{1}=
\left(%
\begin{array}{ccc}
  0 & 1 & 0 \\
  1 & 0  & 0 \\
  0 & 0  & 0 \\
\end{array}%
\right), \ H_{2}=
\left(%
\begin{array}{ccc}
  0 & -i & 0 \\
  i & 0  & 0 \\
  0 & 0  & 0 \\
\end{array}%
\right),\ \ \ \ \ \ \ \ \ \ \ \ \ \ \ \nonumber \end{equation}

\begin{equation}
\ H_{3}=
\left(%
\begin{array}{ccc}
  0 & 0 & 1 \\
  0 & 0  & 0 \\
  1 & 0  & 0 \\
\end{array}%
\right), \ \ \ H_{4}=
\left(%
\begin{array}{ccc}
  0 & 0 & -i \\
  0 & 0  & 0 \\
  i & 0  & 0 \\
\end{array}%
\right).
\end{equation}

For simplicity, we assume $f_{m}(\theta_{m})=f(\vartheta)$ for all $m=1,2,3,4$ and $f_{0}(\theta_{0})=f_{0}(\vartheta)$. After we sample the uncertainty parameters, every sample can be described as follows:
\begin{small}
\begin{equation}\label{general3level}
\left(%
\begin{array}{c}
  \dot{c_{1}}(t) \\
  \dot{c_{2}}(t) \\
  \dot{c_{3}}(t) \\
\end{array}%
\right)=
\left(%
\begin{array}{ccc}
  -1.5f_{0}(\vartheta) i & G_{1}(\vartheta)  & G_{2}(\vartheta) \\
  G^{*}_{1}(\vartheta) & -f_{0}(\vartheta) i  & 0 \\
  G^{*}_{2}(\vartheta) & 0 & 0 \\
\end{array}%
\right) \left(%
\begin{array}{c}
  c_{1}(t) \\
  c_{2}(t) \\
  c_{3}(t) \\
\end{array}%
\right)
\end{equation}
\end{small}
where $G_{1}(\vartheta)=f(\vartheta)[u_{2}(t)-iu_{1}(t)]$, $G_{2}(\vartheta)=f(\vartheta)[u_{4}(t)-iu_{3}(t)]$ and $\vartheta \in [1-E,
1+E]$ . $E \in [0,1]$ is a given
constant and $G^{*}$ is the complex conjugate of $G$.

To construct an augmented system for the training step of the
SLC design, we choose $N$ training samples
(denoted as $n=1, 2, \ldots, N$) through sampling the uncertainties as follows:
\begin{equation}\label{3level-element1}
\left(%
\begin{array}{c}
  \dot{c}_{1,n}(t) \\
  \dot{c}_{2,n}(t) \\
  \dot{c}_{3,n}(t) \\
\end{array}%
\right)=B_{n}(t)\left(%
\begin{array}{c}
  c_{1,n}(t) \\
  c_{2,n}(t) \\
  c_{3,n}(t) \\
\end{array}%
\right),\end{equation}
\begin{equation}B_{n}(t)=
\left(%
\begin{array}{ccc}
  -1.5f_{0}(\vartheta_{n}) i & G_{1}(\vartheta_{n})  & G_{2}(\vartheta_{n}) \\
  G^{*}_{1}(\vartheta_{n}) & -f_{0}(\vartheta_{n}) i  & 0 \\
  G^{*}_{2}(\vartheta_{n}) & 0 & 0 \\
\end{array}%
\right)\nonumber,
\end{equation}
where $G_{1}(\vartheta_{n})=f(\vartheta_{n})[u_{2}(t)-iu_{1}(t)]$, $G_{2}(\theta_{n})=f(\vartheta_{n})[u_{4}(t)-iu_{3}(t)]$. For simplicity, we assume that $f_{0}(\vartheta)=f(\vartheta)=\vartheta$
have uniform distributions over $[1-E, 1+E]$. Now the
objective is to find a robust control strategy $u(t)=\{u_{m}(t), m=1,2,
3,4\}$ to drive the quantum system from
$|\psi_{0}\rangle=|1\rangle$
(i.e.,
$C_{0}=(1, 0, 0)$)
to $|\psi_{\text{target}}\rangle=\frac{1}{\sqrt{2}}(|2\rangle+|3\rangle)$ (i.e., $C_{\text{target}}=(0, \frac{1}{\sqrt{2}}, \frac{1}{\sqrt{2}})$).

If we write (\ref{3level-element1}) as
$\dot{C}_{n}(t)=B_{n}(t)C_{n}(t)$ ($n=1,2,\ldots, N$), we can
construct the following augmented system
\begin{equation}\label{augmented-equation-5samples3level}
\left(%
\begin{array}{c}
  \dot{C}_{1}(t) \\
  \dot{C}_{2}(t) \\
  \vdots \\
  \dot{C}_{N}(t) \\
\end{array}%
\right)=
\left(%
\begin{array}{cccc}
  B_{1}(t) & 0 & \cdots & 0  \\
  0 & B_{2}(t) & \cdots & 0  \\
  \vdots & \vdots & \ddots & \vdots  \\
  0 & 0 & \cdots & B_{N}(t)  \\
\end{array}%
\right) \left(%
\begin{array}{c}
  C_{1}(t) \\
  C_{2} (t)\\
  \vdots \\
  C_{N} (t)\\
\end{array}%
\right).
\end{equation}
For this augmented system, we use the training step to learn
an optimal control strategy $u(t)$ to maximize the following
performance function
\begin{equation}
J(u)=\frac{1}{N}\sum_{n=1}^{N}\vert \langle
C_{n}(T)|C_{\text{target}}\rangle\vert^{2}.
\end{equation}

Now we employ \emph{Algorithm 1} to find an optimal control strategy
$u^{*}(t)=\{u^{*}_{m}(t), m=1,2,3,4\}$ for the augmented
system. Then we apply the optimal control strategy to additional
samples to evaluate the performance of the control strategy.

\subsection{Numerical results}
For numerical experiments on a $V$-type quantum system,
we use the parameter settings listed as follows: The initial state
$|\psi_{0}\rangle=|1\rangle$
and the target state $|\psi_{\text{target}}\rangle=\frac{1}{\sqrt{2}}(|2\rangle+|3\rangle)$; The end time is $T=5$
and the total time duration $[0,T]$ is equally discretized into
$W=200$ time intervals with each time interval $\Delta
t=(t_{w}-t_{w-1})|_{w=1,2,\ldots,W}=T/W=0.025$; The learning rate is
$\eta_{k}=0.2$; The control
strategy is initialized with $u^{k=0}(t)=\{u^{0}_{m}(t)=\sin t,
m=1,2,3,4 \}$.

First, we assume that there exists only the uncertainty $f_{0}(\vartheta)$
(i.e., $f(\vartheta)\equiv 1$), $E=0.21$ and $f_{0}(\vartheta)$ has a
uniform distribution in the interval $[0.79, 1.21]$. To
construct an augmented system for the training step, we use the
training samples for this $V$-type quantum system defined as follows
\begin{equation}
\left\{ \begin{split}
& f_{0}(\vartheta_{n})=1-0.21+\frac{0.21(2n-1)}{7},\\
& f(\vartheta_{n})=1, \\
\end{split}\right.
\end{equation}
where $n=1,2,\ldots,7$. The training performance for the augmented
system is shown in Fig. 1. It is clear that the learning process
converges to a quite accurate stage very quickly. The optimal control strategy is
demonstrated in Fig. 2, which is compared with the initial one. To
test the optimal control strategy obtained from the training step,
we choose $200$ samples obtained by sampling the uncertainty
$f_{0}(\vartheta)$ according to a uniform distribution and
demonstrate the testing performance in Fig. 3. For the 200 tested
samples, an average fidelity of 0.9999 is achieved.

\begin{figure}\label{fig1}
\centering
\includegraphics[width=3.5in]{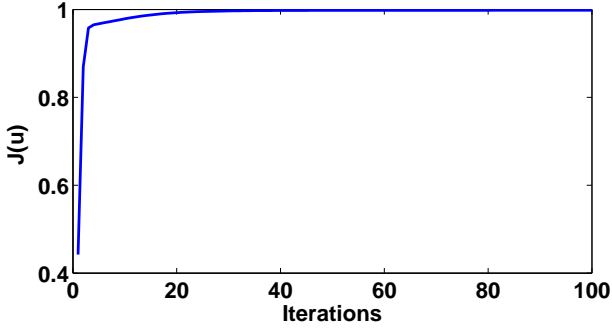}
\caption{Training performance to find the optimal control strategy
by maximizing $J(u)$ for the $V$-type quantum system with only
uncertainty $f_{0}(\vartheta)$ where $f_{0}(\vartheta) \in [0.79,
1.21]$.}
\end{figure}

\begin{figure}\label{fig2}
\centering
\includegraphics[width=3.5in]{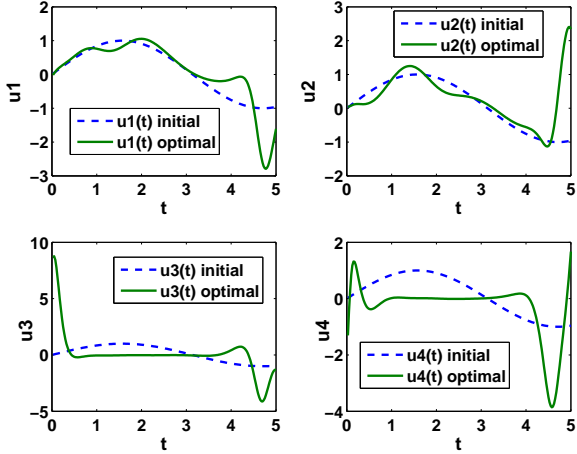}
\caption{The learned optimal control strategy with maximized
$J(u)$ for the $V$-type quantum system with only the uncertainty
$f_{0}(\vartheta)$ where $f_{0}(\vartheta) \in [0.79, 1.21]$.}
\end{figure}

\begin{figure}\label{fig3}
\centering
\includegraphics[width=3.5in]{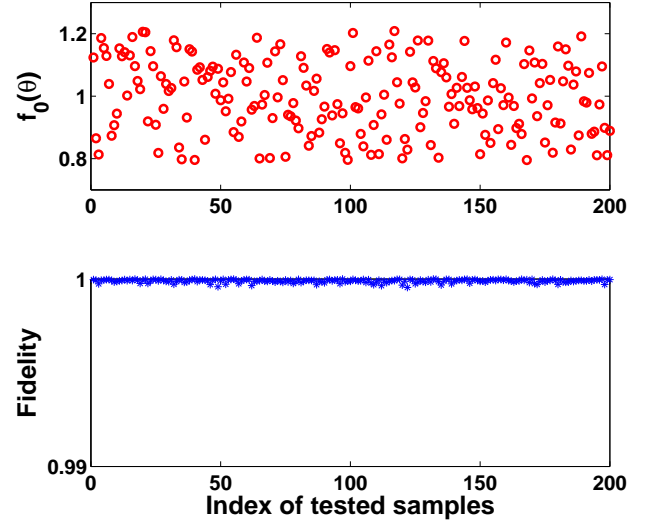}
\caption{The testing performance (with respect to fidelity) of the
learned optimal control strategy for the $V$-type quantum system
with only uncertainty $f_{0}(\vartheta)$ where $f_{0}(\vartheta)
\in [0.79, 1.21]$. For the 200 tested samples, the average
fidelity is 0.9999.}
\end{figure}

Now, we consider the more general case where there exist the
uncertainties  $f_{0}(\vartheta)$ and $f(\vartheta)$. Assume
$E=0.21$, $f_{0}(\vartheta)=1-\vartheta_{0} \cos t$,
$f(\vartheta)=1-\vartheta \cos t$, and both $\vartheta_{0}$ and
$\vartheta$ have uniform distributions on the interval $[-0.21,
0.21]$. To construct an augmented system for the training step, we
use the training samples defined as follows
\begin{equation}
\left\{ \begin{split}
& f_{0}(\vartheta_{n})=1-0.21+\frac{0.21(2\text{fix}(n/7)-1)}{7},\\
& f(\vartheta_{n})=1-0.21+\frac{0.21(2\text{mod}(n,7)-1)}{7}, \\
\end{split}\right.
\end{equation}
where $n=1,2,\ldots,49$, $\text{fix}(x)=\max \{z\in
\mathbb{Z}|z\leq x\}$, $\text{mod}(n,7)=n-7z\ (z\in \mathbb{Z}\
\text{and}\ \frac{n}{7}-1<z\leq \frac{n}{7} )$ and $\mathbb{Z}$ is
the set of integers. The algorithm converges after around 9000
iterations and the optimal control strategy is presented in Fig.
4. To test the optimal control strategy obtained in the training
step, we randomly choose $200$ samples by uniformly sampling
the uncertainties $\vartheta_{0}$ and $\vartheta$ and an average
fidelity of 0.9961 is achieved. However, if we use only one sample (i.e., the nominal system) for training to obtain a control law, the testing performance shows a 0.9152 average fidelity. These numerical results show that
the proposed SLC method using an augmented system for training is
effective for control design of quantum systems with
uncertainties.

\begin{figure}\label{fig4}
\centering
\includegraphics[width=3.5in]{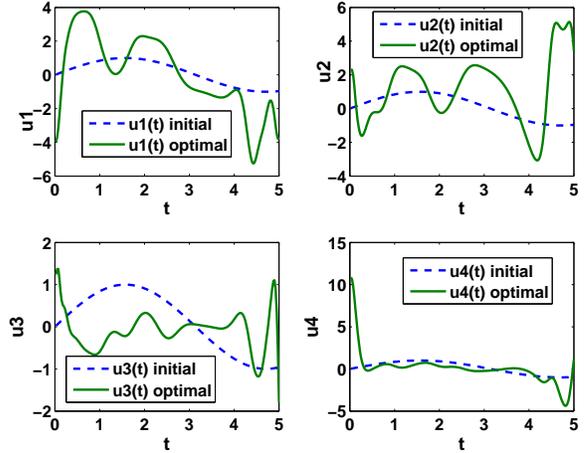}
\caption{The learned optimal control strategy with maximized
$J(u)$ for the $V$-type quantum system with the uncertainties
$f_{0}(\vartheta)$ and $f(\vartheta)$ where
$f_{0}(\vartheta)=1-\vartheta_{0} \cos t$,
$f(\vartheta)=1-\vartheta \cos t$, and both $\vartheta_{0}$ and
$\vartheta$ have uniform distributions on the interval $[-0.21,
0.21]$.}
\end{figure}

\section{Robust entanglement generation in quantum superconducting circuits}\label{Sec5}
Superconducting quantum circuits based on Josephson junctions are
macroscopic circuits which can behave quantum mechanically like
artificial atoms \cite{Clarke and
Wilhelm 2008}-\cite{Dong et al 2015SRep}. These artificial atoms can
be used to test the laws of quantum mechanics on macroscopic
systems and also offer a promising way to realize qubits in
quantum information technology \cite{Makhlin et al 2001}. Superconducting quantum circuits have been recognized as promising candidates for quantum information processing due to their advantages of scalability and design flexibility.
They have been widely investigated
theoretically and experimentally in recent
years \cite{You and Nori 2011}, \cite{Schoelkopf and Girvin 2008}-\cite{Hofheinz et al 2009}.
Superconducting qubits can be controlled by adjusting external
parameters such as currents, voltages and microwave photons, and
the coupling between two superconducting qubits can be turned on
and off at will \cite{Clarke and Wilhelm 2008}. In practical
applications, the existence of noise (including extrinsic and
intrinsic) and inaccuracies (e.g., inaccurate operation in the
coupling between qubits) in superconducting quantum
circuits is unavoidable. In this section, we apply the SLC method to robust control design for a superconducting circuit system with uncertainties.

In
superconducting quantum circuits, two typical classes of qubits are flux qubits
and charge qubits which correspond to different ratios between the Josephson coupling energy
$E_{J}$ and the charging energy $E_{C}$ (For a brief introduction to superconducting quantum circuits, see, e.g., \cite{You and Nori 2005}, \cite{Wendin and Shumeiko 2006}). The simplest charge
qubit is based on a small superconducting island (called a
Cooper-pair box) coupled to the outside world through a weak
Josephson junction and driven by a voltage source through a gate
capacitance within the charge regime (i.e., $E_C\gg E_J$)
\cite{You and Nori 2005}. In practical applications, the Josephson junction in
the charge qubit is usually replaced by a dc superconducting quantum interference device (SQUID) with low
inductance to make it easier to control the qubit. When we
concentrate on a voltage range near a degeneracy point, the Hamiltonian of a superconducting charge qubit can be described as follows
\begin{equation}\label{Hamiltonian-1qubit}
H=F_{z}(V_g)\sigma_{z}-F_{x}(\Phi)\sigma_{x}
\end{equation}
where $F_{z}(V_g)$ is related to the charging energy $E_{C}$ and this
term can be adjusted through external parameters such as the voltage $V_{g}$,
$F_{x}(\Phi)$ corresponds to a controllable term including different
control parameters such as the flux
$\Phi$ in the SQUID, and the Pauli matrices $\sigma=(\sigma_{x},\sigma_{y},\sigma_{z})$ with
\begin{equation}
\sigma_{x}=\begin{pmatrix}
  0 & 1  \\
  1 & 0  \\
\end{pmatrix} , \ \ \ \
\sigma_{y}=\begin{pmatrix}
  0 & -i  \\
  i & 0  \\
\end{pmatrix} , \ \ \ \
\sigma_{z}=\begin{pmatrix}
  1 & 0  \\
  0 & -1  \\
\end{pmatrix} .
\end{equation}

In this section, we consider the coupled two-qubit circuit in
\cite{Wendin and Shumeiko 2006} where an
LC-oscillator is used to couple two charge qubits
(see Fig. 5). Each qubit is realized by a Cooper-pair box with
Josephson coupling energy $E_{Ji}$ and capacitance $C_{Ji}$
($i=1,2$). Each Cooper-pair box is biased by an applied voltage
$V_{i}$ through a gate capacitance $C_{i}$ ($i=1,2$). The
Hamiltonian of the coupled charge qubits can be described as
\cite{Wendin and Shumeiko 2006}
\begin{equation}
\begin{split}
H=&F_{z}(V_{1})\sigma^{(1)}_{z}\otimes I^{(2)}+F_{z}(V_{2})I^{(1)}\otimes \sigma^{(2)}_{z} \\
&-F_{x}(\Phi_1)\sigma^{(1)}_{x}\otimes I^{(2)}-F_{x}(\Phi_2)I^{(1)}\otimes \sigma^{(2)}_{x}\\
&-\chi(t)\sigma^{(1)}_{y}\otimes\sigma^{(2)}_{y}
\end{split}
\end{equation}
where $A^{(j)}$ denotes an operation $A$ on the qubit $j$ and $\otimes$ denotes the tensor product.
Let $u_{1}(t)=F_{z}(V_{1})/\hbar$, $u_{2}(t)=F_{z}(V_{2})/\hbar$, $u_{3}(t)=F_{x}(\Phi_1)/\hbar$, $u_{4}(t)=F_{x}(\Phi_2)/\hbar$, $u_{5}(t)=\chi(t)/\hbar$. For simplicity, we assume the uncertainty parameters $f_{m}(\theta_{m})=\theta_m$ for all $m=1,2,3,4,5$. The Hamiltonian for the practical system can be described as
\begin{equation}\label{coupledHamiltonian2}
\begin{split}
H/\hbar=&\theta_{1}u_{1}(t)\sigma^{(1)}_{z}\otimes I^{(2)}+\theta_{2}u_{2}(t)I^{(1)}\otimes \sigma^{(2)}_{z}\\ & -\theta_{3}u_{3}(t)\sigma^{(1)}_{x}\otimes I^{(2)}-\theta_{4}u_{4}(t)I^{(1)}\otimes \sigma^{(2)}_{x} \\ &-\theta_{5}u_{5}(t)\sigma^{(1)}_{y}\otimes\sigma^{(2)}_{y}
\end{split}
\end{equation}
where $\theta_{m}\in [1-E, 1+E]$ ($m=1,2,3,4,5$).

For practical systems, $E_{J}$ could be around
$10$ GHz and $E_{C}$ could be around
$100$ GHz (e.g., the experiment in \cite{Pashkin
et al 2003} used $E_{J1}=13.4\ \text{GHz}$, and
$E_{C2}=152\ \text{GHz}$). In (\ref{coupledHamiltonian2}), we assume
$u_{1}(t) \in [0, 50.2]\ \text{GHz}$, $u_{2}(t) \in [0,
50.2]\ \text{GHz}$, $u_{3}(t) \in [0, 11.1]\ \text{GHz}$, $u_{4}(t)
\in [0, 11.1]\ \text{GHz}$, $|u_{5}(t)|\leq 0.5\ \text{GHz}$, and the
operation time $T=2\ \text{ns}$. As an example, we let the task be
to generate a maximally entangled state
$|\psi_{\text{target}}\rangle=\frac{1}{\sqrt{2}}(|g_{1},g_{2}\rangle+|e_{1},e_{2}\rangle)$,
where $|g_{j}\rangle$ and $|e_{j}\rangle$ denote the ground state
and the excited state of qubit $j$, respectively. In quantum
information, we usually use $|0\rangle$ (or $|1\rangle$) to denote
$|g\rangle$ (or $|e\rangle$).

Quantum entanglement is a unique quantum phenomenon that occurs
when quantum subsystems are generated or interact in ways such
that the quantum state of each subsystem cannot be described
independently \cite{Nielsen and Chuang 2000}, \cite{Cui et al
2009}, \cite{Mohamed et al 2014IFAC}. Quantum entanglement shows
nonclassical correlation and has been demonstrated as an important
physical resource in quantum cryptography, quantum communication
and quantum computation \cite{Nielsen and Chuang 2000},
\cite{Pan1997}. We may use concurrence to measure how entangled a
two-qubit state is \cite{Wootters 1998}. For a two-qubit state
$\rho$, let $\rho^{*}$ denote the complex conjugate of $\rho$,
$\tilde{\rho}=(\sigma_{y}\otimes \sigma_{y})\rho^{*} (\sigma_{y}\otimes \sigma_{y})$ and $R=\sqrt{\sqrt{\rho}\tilde{\rho}\sqrt{\rho}}$. Let $\lambda_{1},\lambda_{2},\lambda_{3},\lambda_{4}$ be the eigenvalues of $R$ in decreasing order. The concurrence is defined as $$\mathcal{C}(\rho)\equiv\max (0, \lambda_{1}-\lambda_{2}-\lambda_{3}-\lambda_{4}).$$

Let $\theta_{1}=\theta_{2}$, $\theta_{3}=\theta_{4}$,
$\theta_{j}\in [0.79, 1.21]$. In the training step, we uniformly select 7 values for each uncertainty parameter to generate samples for constructing an augmented
system. In the testing step, we assume $\theta_{j}$ has a
truncated Gaussian distribution. We also assume $|\psi_{0}\rangle=|g_{1},g_{2}\rangle$.

In numerical experiments, we divide $t\in[0, 2]\ \text{ns}$ equally
into $200$ time intervals. The control fields are initialized as:
$u_{1}(0)=u_{2}(0)=u_{3}(0)=u_{4}(0)=\sin t+5 \ \text{GHz}$,
$u_{5}(0)=0.25\sin t \ \text{GHz}$. Assume the control fields satisfy
$|u_{j}(t)|\leq V$. During the learning process, if the calculated
control $u_{j}^{k}\geq V$ using the gradient algorithm, we let
$u_{j}^{k}= V$. Similarly, if the calculated control
$u_{j}^{k}\leq -V$, we let $u_{j}^{k}= -V$. The learning algorithm
converges after about $9800$ iterations and the performance is
shown in Fig. 6. A set of optimal control fields is shown in Fig.
7. Then the learned optimal control fields are applied to 2000
samples that are generated through selecting different values of
uncertainty parameters according to the truncated Gaussian
distribution with mean $\mu=1$ and standard
deviation $\nu=0.07$ within the interval of $[1-3\nu,
1+3\nu]=[0.79, 1.21]$. The average concurrence is obtained from
2000 samples. When $\theta_{i}\in [0.79, 1.21]$, the testing
process (using 2000 randomly selected samples) shows that the
average fidelity is 0.9992 and the average concurrence is 0.9981.
Hence, the learned control fields can still drive the system to a
maximally entangled state with high concurrence when the
uncertainty parameters have $42\%$ fluctuations.

\begin{figure}
\centering
\includegraphics[width=3in]{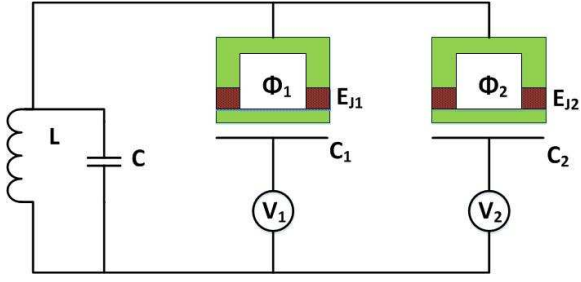}
\caption{Two charge qubits coupled to a common LC-oscillator.}
\end{figure}

\begin{figure}
\centering
\includegraphics[width=3.5in]{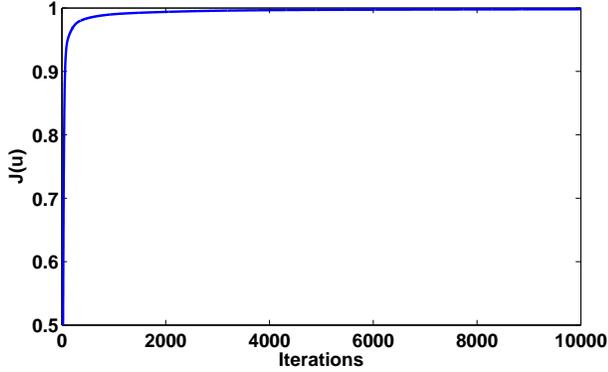}
\caption{Training performance to find the optimal control strategy
by maximizing $J(u)$ for coupled qubits via an LC-oscillator when
the uncertainty bound is $E=0.21$.}
\end{figure}

\begin{figure}
\centering
\includegraphics[width=3.5in]{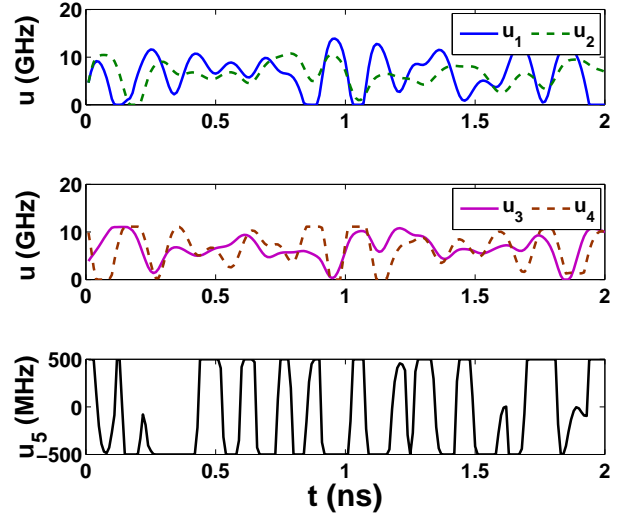}
\caption{The learned control field for the problem of coupled
qubits via an LC-oscillator when the uncertainty bound is $E=0.21$.}
\end{figure}

\section{Robust entanglement control between two atoms in a cavity}\label{Sec6}
\subsection{The system}
In this section, we apply the SLC method to a quantum system consisting of two
two-level atoms interacting with a quantized field in an optical cavity (see Fig. 8) or in a microwave cavity. This model has been wieldy used in experimental quantum optics and quantum information \cite{Altomare2010a,Mabrok et al 2014CDC,Ficek2002,Majer2007,Filipp2011}.  The two-qubit system  can be
represented using four basis vectors $|g_{1},g_{2}\rangle$, $|e_{1},g_{2}\rangle$, $|g_{1},e_{2} \rangle$ and $|e_{1},e_{2}\rangle$
where $|e_j\rangle$ denotes the excited state of the atom
$j$ and $|g_j\rangle$ denotes the ground state of the atom $j$. The Hamiltonian which describes the two-atom system interacting with a quantized field can be written as follows
\begin{equation}
H(t)=H_{0}+H_{I}+H_{u}.  \label{ham1}
\end{equation}
Here, the first term in \eqref{ham1}
\begin{equation}
H_{0}=\frac{1}{2}\sum\limits_{i=1}^{2}\omega _{Ai}\sigma _{z}^{(i)} +\omega_r a^\dag a
\end{equation}
is the Hamiltonian describing the energy of the
atoms and the quantized field.  $\omega_{Ai}$ is the  atomic transition frequency for
atom $i$, and the  operators $a,a^\dag$ represent the annihilation and creation operators. An annihilation operator lowers the number of particles in a given state by one. A creation operator is the adjoint of the annihilation operator.
The second term
\begin{equation}\label{hamI}
    H_{I}=\sum\limits_{i\neq j}^{2}\Omega _{ij}\sigma _{+}^{(i)}\otimes \sigma
_{-}^{(j)}+\sum\limits_{ j}^{2}\nu_j (a^\dag \sigma
_{-}^{(j)}+a \sigma
_{+}^{(j)}),
\end{equation}
represents the interactions. The first term is the dipole-dipole interaction between the two qubits
(atoms), where $\Omega_{ij}$ is the dipole-dipole interaction
parameter, $\sigma
_{+}^{(i)}=|e_{i}\rangle \langle g_{i}|$ and $\sigma
_{-}^{(i)}=|g_{i}\rangle \langle e_{i}|$. Most atoms have  attractive forces
between each other due to fluctuation dipole moments when the electrons of an
atom leave the positively charged nucleus unshielded. This is called
dipole-dipole interaction \cite{Jaksch2000}. The second term in the Hamiltonian in \eqref{hamI} represents the interaction between the field and the atoms, where $\nu_j$ is the coupling constant between the atoms and the quantized field.
The last term in the Hamiltonian given in \eqref{ham1} is the control Hamiltonian:
 \begin{equation*}
 \begin{split}
H_{u} = & \sum\limits_{i=1}^{2}u_{\omega _{Ai}}\sigma _{z}^{(i)} +u_{\omega_r} a^\dag a\notag+\sum\limits_{i\neq j}^{2}u_{\Omega _{ij}}\sigma _{+}^{(i)}\otimes \sigma
_{-}^{(j)}\\
&+\sum\limits_{ j}^{2}u_{\nu_j} (a^\dag \sigma
_{-}^{(j)}+a \sigma
_{+}^{(j)}).
\end{split}
\end{equation*}

The aim is to find functions $u_{\omega _{Ai}},u_{\omega_r},u_{\Omega _{ij}},u_{\nu_j}$ to drive the quantum system to
a particular target state with a desired level of fidelity even when uncertainties exist.
The proposed control Hamiltonian includes  several terms. The first two
terms $u_{\omega _{Ai}}\sigma _{z}^{(i)} +u_{\omega_r} a^\dag a$ represent the control of the energy in the system through the atomic transition frequency  $\omega_{Ai}$ and field frequency $\omega_r$. The term $u_{\Omega _{ij}}\sigma _{+}^{(i)}\otimes \sigma
_{-}^{(j)}$ represents the change in the
dipole-diploe interaction between the atoms, and can be
controlled by changing the distance between the two atoms, or
tuning the frequency of the driving field. The term  $u_{\nu_j} (a^\dag \sigma
_{-}^{(j)}+a \sigma
_{+}^{(j)})$ represents the control of the interaction between the atoms and the field. In this paper, we let $u_{1}(t)=u_{\omega _{A1}}=u_{\omega _{A2}}$, $u_{2}(t)=u_{\omega_r}$, $u_{3}(t)=u_{\Omega _{ij}}$, $u_{4}(t)=u_{\nu_1}$ and $u_{5}(t)=u_{\nu_2}$.

We consider that a steady number of photons is in the cavity. The state of the quantum system consisting of two atoms interacting with a quantized field in a cavity can be described as follows
\begin{align*}
|\psi(t)\rangle &=c_{1}(t)|n+2,g_{1},g_{2}\rangle +c_{2}(t)|n+1,e_{1},g_{2}\rangle \\
&
+c_{3}(t)|n+1,g_{1},e_{2}\rangle +c_{4}(t)|n,e_{1},e_{2}\rangle,
\end{align*}
where $|n\rangle$ is the number state of photons in the cavity and
the complex coefficients $c_{1}(t),c_{2}(t),c_{3}(t)$ and $c_{4}(t)$ satisfy
$
|c_{1}(t)|^{2}+|c_{2}(t)|^{2}+|c_{3}(t)|^{2}+|c_{4}(t)|^{2}=1$.

\begin{figure}
\centering
\includegraphics[width=3.5in]{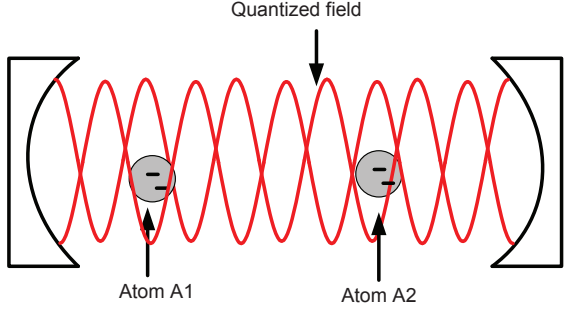}
\caption{Schematic for two atoms interacting with a quantized field in a cavity.}
\end{figure}

We assume that
the Hamiltonian with uncertainties can be written as follows:
\begin{equation}
H(t)=\theta_{0}H_{0}+ \theta_{I}H_{I}+\theta_{u} H_{u},
\label{ham2}
\end{equation}
where $\theta_{0}$, $\theta_{I}$ and $\theta_{u}$ represent
uncertainty parameters in the free Hamiltonian, the interaction Hamiltonian
and the control Hamiltonian, respectively. We assume that the
uncertain parameters satisfy $\theta_{0} \in [1-E, 1+E]$, $\theta_{I}\in [1-E, 1+E]$ and $\theta_{u}\in [1-E, 1+E]$.

The density matrix of the system under consideration is given as follows:
 \begin{align}\label{roo1}
   \rho(t)=|\psi(t)\rangle \langle\psi(t)|.
 \end{align}
The density matrix carries information about the two subsystems of the  atoms and the field. However, we are interested in the entanglement between the two atoms. Hence, the field needs to  be  traced out of the  density matrix \eqref{roo1}. This task can be accomplished by a partial trace operation $\text{Tr}_{f}$ over the field (see, e.g., \cite{Nielsen and Chuang 2000} for a detailed description of the partial trace).

Now, we define  the performance function  as follows
\begin{equation}
J(u)=\text{Tr}[\sqrt{\sqrt{\rho_{A}(T)} \rho^{A}_{\text{target}} \sqrt{\rho_{A}(T)}}]\label{perfor}
\end{equation}
where $\rho_{A}(T)=\text{Tr}_f[|\psi(T)\rangle \langle\psi(T)|]$ is the density matrix  of the two-atom subsystem at the end time $T$ of the
evolution, and $\rho^{A}_{\text{target}}=|\psi^{A}_\text{target}\rangle \langle\psi^{A}_\text{target}|$ is the target density matrix of the two-atom subsystem. During the learning process, this
performance function is used to measure the fidelity of the
system for a given control law. An optimal control law can be found by maximizing $J(u)$.

\subsection{Numerical results}\label{numerical_res}
For the proposed two-qubit system (two atoms) interacting with a quantized electromagnetic field, we are interested  in generating maximum entanglement $|\psi^{A}_{\text{target}}\rangle=\frac{1}{\sqrt{2}}(|e_1,g_2\rangle+|g_1,e_2\rangle)$ between the two
qubits (atoms). The parameters in atomic units are set as follows: The atomic transition
frequencies are $(\omega_1,\omega_2)=( 6.44,3.34)$ and the dipole-dipole
interaction parameter is $\Omega_{12}=0.0259$. The same relative relationship
between the atomic transition frequencies and the dipole-dipole interaction as that in the experiment in
\cite{Majer2007} has been considered. The evaluation time is
$T=2$ and the interval $[0,T]$ is discretized equally into $W=350$
time steps, where $\Delta t=\frac{T}{W}$. The learning rate is
set as $\eta_{k}=0.1$. The initial control law is assumed to be  $u^{0}_i=\sin
t$ ($i=1,2,\cdots,5$).

\begin{figure}
  \centering\includegraphics[width=8.8cm]{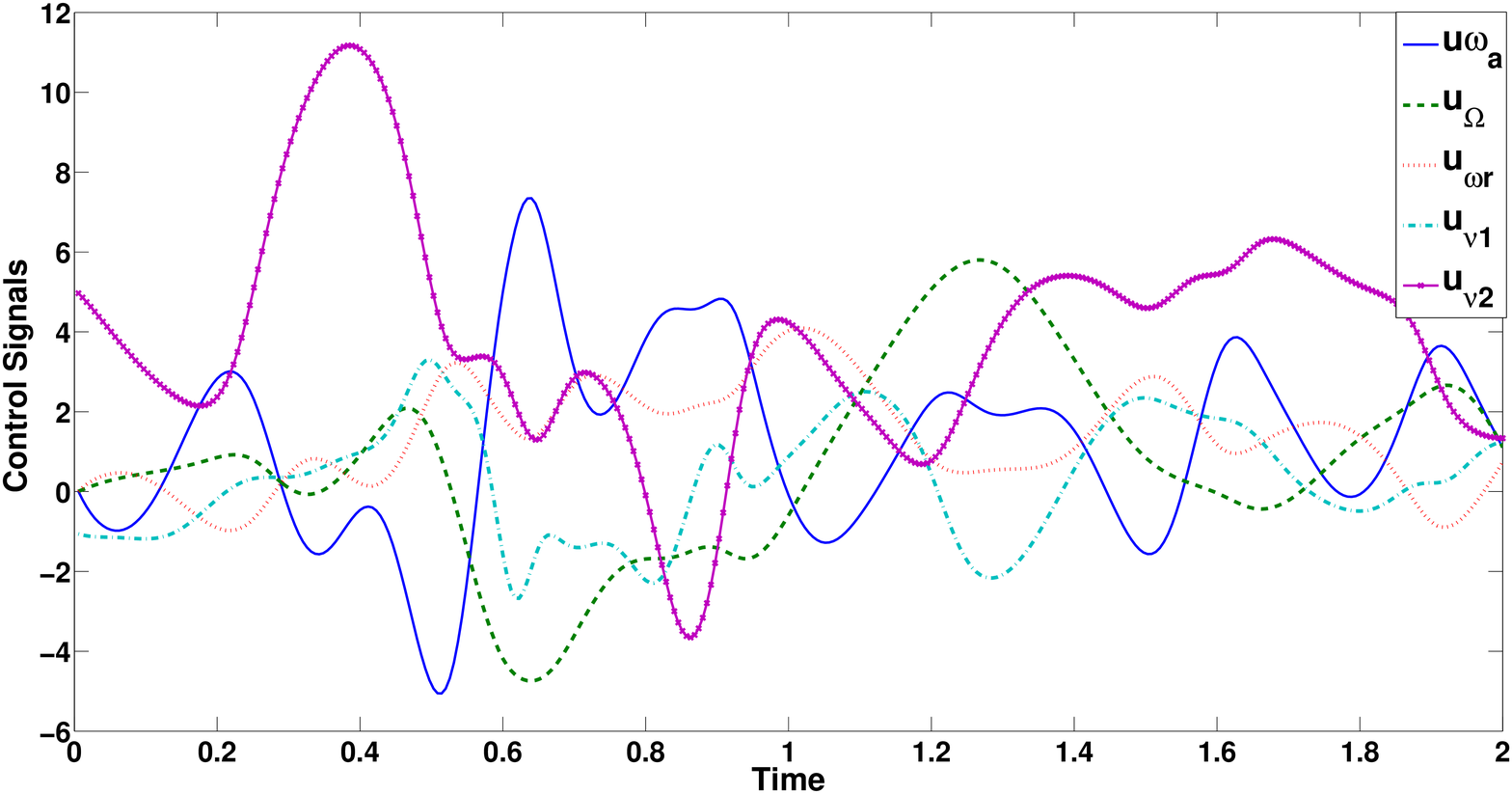}\\
  \caption{The learned control strategy, where the  initial state is
$|\psi^{A}(0)\rangle=|g_1,g_2\rangle$ and
the target state
$|\psi^{A}_{\text{target}}\rangle=\frac{1}{\sqrt{2}}(|e_1,g_2\rangle+|g_1,e_2\rangle)$. The uncertainty bound is $E=0.2$, $u_{1}(t)=u{\omega_a}$, $u_{2}(t)=u_{\omega r}$, $u_{3}(t)=u_{\Omega}$, $u_{4}(t)=u_{\nu 1}$ and $u_{5}(t)=u_{\nu 2}$.
}\label{testing_all_1}
\end{figure}

The uncertainty parameters $\theta_0$, $\theta_I$ and $\theta_u$ are assumed to have a uniform
distribution in the interval $[1-E,1+E]$ with $E=0.2$. We select 5 values for each uncertainty parameter to construct an augmented system.
We assume that the five controls $u_{m}(t)$ ($m=1,2,3,4,5$) are permitted in the Hamiltonian $H(t)$. The initial state is chosen to be the ground state
$|\psi^{A}(0)\rangle=|g_1,g_2\rangle$ and
the target state is chosen to be
$|\psi\rangle=\frac{1}{\sqrt{2}}(|e_1,g_2\rangle+|g_1,e_2\rangle)$. The algorithm converges with around 8000 iterations. In the evaluation step, we select 500 additional samples to test the
control performance. The average fidelity we can achieve is 0.9966 and the average concurrence is 0.9880. Fig. \ref{testing_all_1} gives the learned control strategy $u_{m}(t)$ ($m=1,2,3,4,5$).

\section{Conclusion}\label{Sec7}
In this paper, we presented a systematic numerical methodology for robust control
design of quantum systems. The proposed sampling-based learning control (SLC) method
includes two steps of ``training" and ``testing".
In the training step, the control is learned using a gradient flow based
learning algorithm for an augmented system
constructed from samples. The learned control
is evaluated for additional samples in the testing step. The proposed numerical method has been applied to three significant examples of quantum robust control including state preparation in a three-level system, entanglement generation in a superconducting quantum circuit and in a two-atom system interacting with a quantized field in a cavity. In these examples, we considered the uncertainty parameters to have uniform distributions, truncated Gaussian distributions or possibly time-varying distributions. However, numerical results showed that the uniform distribution is a sound choice for sampling the uncertainty parameters in the training step. Before we start the training step, we may first analyze the controllability of the nominal system (e.g., using Lie group and Lie algebra theory \cite{Dong and Petersen 2010IET}, \cite{D'Alessandro 2007}). Such an analysis may be difficult when we consider the constraints on control strengths and control durations. Even if we prove that the nominal system is not controllable, it may still be possible to achieve accurate state transfer between specific states that correspond to some useful practical applications. One advantage of the proposed method is that it is numerically tractable in achieving convergence since we can use a small number of samples (e.g., five or seven) for each uncertainty parameter in the training step to obtain excellent performance. Based on the training performance (whether the cost $J$ is close to one), it is easy to verify when an optimal solution has been found. The results in this paper have demonstrated the effectiveness of the SLC method for
control design of quantum systems even when the uncertainty parameters have quite large fluctuations.

\section*{Acknowledgment}
D. D and C. C. would like to thank Ruixing Long for helpful discussions.

\end{document}